\documentclass[aps,12pt,preprint,amssymb,tightenlines,titlepage
,showpacs%
,prd%
,nobibnotes%
,nofootinbib%,superscriptaddress
]{revtex4}
\usepackage[dvips]{graphicx}
\def\ga{\mathrel{\raise.3ex\hbox{$>$\kern-.75em\lower1ex\hbox{$\sim$}}}}
\def\la{\mathrel{\raise.3ex\hbox{$<$\kern-.75em\lower1ex\hbox{$\sim$}}}}
\newcommand{\beq}{\begin{equation}} 
\newcommand{\eeq}{\end{equation}} 
\newcommand{\bea}{\begin{eqnarray}} 
\newcommand{\eea}{\end{eqnarray}}
\begin{document}
\preprint{\parbox{7cm}{{\tt astro-ph/0311293}\\
                           TPI--MINN--03/31}} 
\vspace*{2cm} 

\title{Phantom Cosmologies}

\author{\bf Vinod B. Johri}
\altaffiliation[Permanent address: ]{Department of Mathematics and Astronomy, 
Lucknow University,\\ Lucknow 226007, India. E-mail:vinodjohri@hotmail.com }
\affiliation{William I. Fine Theoretical Physics Institute,
University of Minnesota, Minneapolis, MN 55455, USA}

%%%%%%%%%%%%%%%%%%%%%%%%% 
\begin{abstract}
The dynamics of a minimally coupled scalar field in the 
expanding universe is discussed with special reference to phantom cosmology.
The evolution of the universe with a phantom field vis-a-vis a quintessence 
field is compared. Phantom cosmologies are found to have two special features
i) occurrence of a singularity where the scale factor, the energy density and 
Ricci curvature scalar diverge to infinity. 
This singularity occurs at a finite timexs, 
depending on the value of  $w$ during cosmic evolution, ii) degeneracy in the 
determination of $w(z_m)$ for a given  transition redshift $z_m$
which seems to impart similar observational properties to corresponding 
phantom and quintessence models and makes both of them compatible with the
 cosmological observations. 
Although due to the uncertainties in the measurement of 
the Hubble constant $H_0$, the Hubble dependent observational parameters 
 yield only loose constraints over the range of $w$, 
the duality in the determination of $w$ with respect to transition redshift
may be used to constrain $w$.
 An observational test, 
based upon the observations of low redshift galactic clusters, is suggested 
to  discriminate between the quintessence and phantom dark energy.
\end{abstract} 

\maketitle

\baselineskip=18pt

\section{ INTRODUCTION }
The combined analysis of SNe Ia observations~\cite{1,2}, 
galaxy cluster measurements~\cite{3}
 and the latest CMB data~\cite{4} provides compelling evidence for 
the existence of dark energy
which dominates the present day universe and accelerates the cosmic expansion. 
The recent detection of Integrated Sachs-Wolfe effect~\cite{5} 
also gives a strong and
independent support to dark energy. In principle, any physical 
field with positive energy
density and negative pressure, which violates the strong energy 
condition, may cause the
dark energy effect of repulsive gravitation. 
Of late, phantom fields~\cite{6}
have emerged as
 potential candidates for dark energy. Scalar fields with super-negative 
equation of state
$(p=w\rho  , \,w< -1)$ are called phantom fields as their energy density 
increases with the
expansion of the universe in contrast to 
quintessence energy density $(w> -1)$  which 
scales down with the cosmic expansion. The phantom models violate the dominant 
energy condition $( p+\rho) < 0$ as such they may not be 
physically stable models of dark energy; but, strangely enough, 
phantom energy is found to be compatible with most of the classical 
tests of cosmology~\cite{6} based on current data from SNe Ia observations, 
CMB anisotropy and mass power spectrum.

The peculiar nature of phantom energy, violation of dominant 
energy condition and its
strange consequences, possible rip-off of the large and 
small scale structures of matter, 
occurrence of future singularity and probable decay of phantom e
nergy have attracted many 
cosmologists~\cite{7}-\cite{24},~\cite{52,53,54} and made `phantom cosmologies' 
a hot topic of research.

In section 2, we have discussed the dynamics of  minimally coupled 
scalar fields with 
special  reference to phantom fields. There is extensive literature 
in cosmology on
rolling scalar fields~\cite{22,23}, quintessence fields and tracker 
fields~\cite{24}-\cite{40},~\cite{50,51}, the 
cosmological constant $\Lambda$~\cite{25,41,42,49} and other forms of
 dark energy. The major
problem in cosmology is to identify the form of dark energy that dominates the 
universe today whether it is phantom energy, quintessence, simply $\Lambda$ or 
something else. Maor, Brustein and Steinhardt~\cite{43} 
have discussed the degeneracy
 in the measurement of the dark energy parameter $w$ from SNe 
Ia data, its time 
variation and pitfalls in taking $w$ to be a constant. 

In section 3, we have computed the present age $ t_0 $ of the universe 
 in two steps.
First we calculate the expansion age up to the end of matter-dominated
 era, denoted by
$t_m$. In order to supplement it with the expansion age during 
dark energy dominated era, 
we express $t_0$ in terms of $t_m$ and thereby we calculate $t_0$. The 
advantage of this method is that the expansion age in the 
two segments can be expressed 
separately in terms of the redshift $z_m$  at the end of
 matter-dominated era which is 
again a function of the parameter $w$. 

In section 4, assuming $w$ to be constant, we have  
discussed a kind of degeneracy  in the
value of $w(z_m)$ which leads to duality in the behavior of 
phantom and quintessence
models with respect to transition redshift from deceleration to 
accelerating phase of expansion.
In fact two distinct values of parameter $w$, 
usually one lying in the range of quintessence
field and another in the range of phantom field, 
lead to the same transition redshift $z_m$.

In section 5, we have tried to constrain the range of the 
dark energy parameter $w$ on the basis 
of data analysis of Kiselev~\cite{44}, Freedman and 
Turner~\cite{45} Schubnell~\cite{46} and  the precise 
observational data  from WMAP~\cite{47} in combination with SDSS~\cite{48}. 
In section 6, we conclude with some remarks on phantom energy.

\section{DYNAMICS OF PHANTOM COSMOLOGY }

Consider a 2-component cosmic fluid in a Friedmann universe comprising 
(i) pressure-free matter of energy density $\rho_m$ 
and (ii) a minimally coupled scalar field
of energy density $\rho_x$ and equation of state $ p=w\rho $ which contributes 
to dark energy in the universe.The energy densities $\rho_m\sim a^{-3}$ and 
$\rho_x\sim a^{-3(1+w)}$ evolve independently in the expanding universe.

 The dark energy might be due to 

   (a) quintessence field if $-1<w<-\frac{1}{3}$ 

   (b) cosmological constant $\Lambda$ if $w= -1$ 

or (c) phantom field if $w<-1$.

In the above classification, the equation of state parameter $w$ plays 
the role of dark energy parameter. 

The Friedmann equations are

%%%%%%%%%%%%%%%%%%%%%%%%%%%%%%%%%%%%%%%%%%%%%%%%%%%%%%%%%%%
\begin{equation}
\frac{{\dot a}^{2}}{a^2}\,=\,
\frac{8\pi G}{3} \bigl[ \rho_m+\rho_x \bigr] 
  \,=\,H_0^2 \bigl[ \Omega_m^0(a_0/a)^{3} +\Omega_x^0(a_0/a)^{3(1+w)}    \bigr]
\end{equation}
%%%%%%%%%%%%%%%%%%%%%
and
\bea
\frac{\ddot a}{a} &=&-\frac{4\pi G}{3}[\rho_m+\rho_x(1+3w)]
=\,-\frac{4\pi G}{3}\, \rho_x \, [\Omega_x^{-1}+3w]  \nonumber\\
&=&-\frac{4\pi G}{3}\, \rho_x\, 
           \left[ \frac{\Omega_m^0}{\Omega_x^0}
  \left( \frac{a_0}{a} \right)^{-3w}+1+3w \right]
\eea
The cosmic expansion decelerates as long as
\begin{equation}
\Omega_x^{-1} + 3w > 0
\end{equation}
With the growth of phantom energy density parameter, 
$\Omega_x^{-1} $ goes on decreasing
with time until the transition to accelerating phase 
takes place at cosmic time $t = t_m$. The
transition epoch $t_m$ corresponds to the red-shift $z_m$ given by
%%%%%%%%%%%%%%%%%%%%%%%%%%%%%%%%%%%%%%%%%%%%
\begin{equation} 
1+z_m\,=\, \Bigl[ \frac{-(3w+1)\Omega_x^0}{\Omega_m^0} \Bigr]^{-\frac{1}{3w}}
\end{equation}

It may be emphasized that the transition epoch $t_m$ 
marks the end of the 'effective matter
dominated' era or the beginning of the accelerating phase 
in the cosmic expansion after
which the large scale structure formation in the 
universe must cease although $\Omega_m$
may still be greater than $\Omega_x$. It is
 evident from Eq.(4) that the transition epoch
$t_m$ depends upon the choice of $w$.   
FIG~\ref{fig3} shows the variation of $z_m$ with $w$. 

%%%%%%%%%%%%%%%%%%%%%%%%%% Figure 1 %%%%%%%%%%%%%%%%%%%%%%%%%%%%%%%% 
\begin{figure}[t]  
\centering\includegraphics[height=8cm]{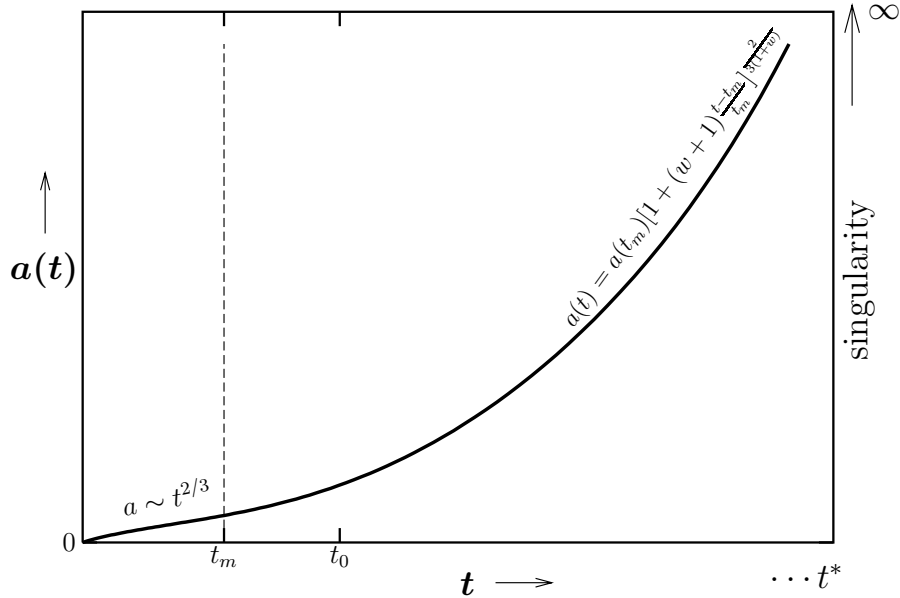}

\caption[]{\it  Expansion of the universe with matter and  
phantom fields: up to
the end of the matter dominated phase $t_m$, the universe 
undergoes Einstein-de Sitter
expansion with deceleration. For $t>t_m$ the cosmic expansion accelerates.
In the case of phantom fields ( $w<-1$), the scale factor diverges to
infinity at finite time $t=t^*$. }
\label{fig1}  
\end{figure}  
%%%%%%%%%%%%%%%%%%%%%%%%%%%%%%%%%%%%%%%%%%%%%%%%%%%%%%%%%%%%%%%%%%
 
As $\Omega^{-1}_x$ decreases further after the transition epoch
%%%%%%%%%%%%%%%%%%%%%%%%%%%%%%%%%%%%%%%%%

\begin{equation}
\Omega^{-1}_x + 3w\,<\,0
\end{equation} 

and the Hubble expansion in the accelerating phase of 
the universe is given by Eq.(1)
%%%%%%%%%%%%%%%%%%%%%%%%%%%%%%%%%%%%%%%%%%%%%%%%%%

%%%%%%%%%%%%%%%%%%%%%%%%%%%%%%%%%%%%%%%%%%%%%%%%%%

\begin{equation}
\frac{\dot a}{a}\,=\,H_0\sqrt{\Omega_x^0} \left( \frac{a_0}{a} \right)^{3(1+w)/2}
      \left[ 1+\frac{\Omega_m^0}{\Omega_x^0}
     \left( \frac{a}{a_0} \right)^{3w} \right]^{\frac{1}{2}}
\end{equation}

Expanding binomially and integrating, we get 
%%%%%%%%%%%%%%%%%%%%%%%%%%%%%%%%%%%%%%%%%%%%
\begin{equation}
\chi t +c=\left( \frac{a}{a_0} \right)^{3(1+w)/2}\left[\frac{2}{3(1+w)}
      -\frac{\Omega_m^0(a)^{3w}}{\Omega_x^0(a_0)^{3w}(9w+3)}+ ...\right]
\end{equation}
where $\chi = H_0 \sqrt{\Omega_x^0}$.
The binomial expansion of the right hand side of Eq.(6) is
valid under the condition 
$\frac{\Omega_m^0}{\Omega_x^0}(\frac{a}{a_0})^{3w} <1<-(3w+1)$
which holds  during the accelerating phase 
over the range (i)$a <a_0$, $w<0$ and 
$\Omega_m^0<\Omega_x^0$ and (ii) $a<a_0$,\,$w<0$ 
and $(1+z)^{-3w}<-\frac{(3w+1)\Omega_x^0}{\Omega_m^0}$. This ensures that the
accelerated expansion during the regime $\Omega_m >\Omega_x$ 
continues as long as 
$-(3w+1)\Omega_x^0 >\Omega_m^0$.

Since $w<0$, the successive terms on the right hand side 
of Eq.(7) decrease by O(3w) of magnitude and the scale
factor is effectively given by
%%%%%%%%%%%%%%%%%%%%%%%%%%%%%%%%%%%%%%%%%%%%%%%%%%%
\begin{equation}
a^{3(1+w)/2}(t) \,=\, \frac{3(1+w)}{2} \chi t + c
\end{equation}

It shows that the contribution of the 
matter density is almost negligible during phantom dominated
universe since the contribution of $\Omega_m^0$
 falls down steeply by 3 orders of magnitude or
more with each successive term in Eq.(7). 
According to Eq.(1)also, the cosmic expansion in the
accelerating phase is essentially driven by the phantom 
field since its energy density scales up as 
$\rho_x\sim a^{-3(1+w)}$ whereas $\rho_m$ scales down as $\sim a^{-3}$.

During the deceleration phase, the Hubble expansion is given by
%%%%%%%%%%%%%%%%%%%%%%%%%%%%%%%%%%%%%%%%%%%%%%%%
\begin{equation}
\frac{\dot a}{a} \,=\,H_0\sqrt{\Omega_m^0} \left( \frac{a_0}{a}\right)^{3/2}
\left[1+\frac{\Omega_x^0}{\Omega_m^0}\left(\frac{a_0}{a}\right)^{3w}\right]^{1/2}
\end{equation}

The deceleration condition (3) implies that 
$\frac{\Omega_x^0}{\Omega_m^0}(\frac{a_0}{a})^{3w}
< -\frac{1}{3w+1}\, <1$ since $w<-1$.

Therefore expanding   Eq.(9) binomially and integrating  we get
%%%%%%%%%%%%%%%%%%%%%%%%%%%%%%%%%%%%%%%%%%%%%%%%%%%%%%%%
\begin{equation}
\xi t \,=\,{a}^{3/2}\left[\frac{2}{3}-\frac{{(1+z)^{3w}}\Omega_x^0}{3(1-2w)\Omega_m^0}
       + \frac{(1+z)^{6w}}{4(1-4w)}
 \left(\frac{\Omega_x^0}{\Omega_m^0}\right)^2 + ... \right]
\end{equation}

Since $w<0$, the first term on the right hand side 
of Eq.(10) dominates while the remaining terms 
decrease for high redshifts. Therefore during the matter dominated era, 
the scale factor is given by

\begin{equation}
%%%%%%%%%%%%%%%%%%%%%%%%%%%%%%%%%%%%%%%%%%%%%%%%%%%%%5
a^{3/2}(t) \,=\, \frac{3}{2}\xi t
\end{equation} 

Eq.(11) holds at the epoch $t=t_m$, as such
%%%%%%%%%%%%%%%%%%%%%%%%%%%%%%%%%%%%%%%%%%%%
\begin{equation}
%%%%%%%%%%%%%%%%%%%%%%%%%%%%%%%%%%%%%%%%%%%%%%%%%%%%%%%%%%%%%%%%%%
a^{3/2}(t_m)\,=\,\frac{3}{2}\xi t_m
\end{equation}

At the beginning of the dark energy dominated phase, Eq.(8) gives
%%%%%%%%%%%%%%%%%%%%%%%%%%%%%%%%%%%%%%%%%%%
\begin{equation}
{a}^{3(1+w)/2}(t_m)\,=\,\frac{3}{2}\,\chi t_m + c
\end{equation}

Matching the junction conditions at $t = t_m$, Eqs,(8),(12) and (13) yield 
%%%%%%%%%%%%%%%%%%%%%%%%%%%%%%%%%%%%%%%%%%%%%%%%%%
\begin{equation}
\left[\frac{a(t)}{a(t_m)}\right]^{3(1+w)/2} 
=\, 1+ \frac{{3/2}(w+1)\chi(t-t_m)}{a^{3(1+w)/2}(t_m)}
     =1+ (w+1)\frac{t - t_m}{t_m}
\end{equation}

Therefore the  scale factor in the phantom (dark energy) 
dominated universe is given by 
%%%%%%%%%%%%%%%%%%%%%%%%%%%%%%%%%%%%%%%%%%%%%%%%%%%%%
\begin{equation}
a(t) \,=\,\frac{a(t_m)}{[-w+(1+w)t/t_m]^{-\frac{2}{3(1+w)}}} \,\,\, for \;\; t >t_m
\end{equation}

Since $1+w<0$ for phantom fields, $a(t)$ diverges to infinity at 
$t^* = \frac{w}{1+w} t_m$. Prior to
blow over time $t^*$, $H>0$ and the deceleration parameter $q = -1 +3(1+w)/2$
in contrast to $q_0 = 1/2 +3/2 w\Omega_x $ at the present epoch. 
On the contrary in the
quintessence dominated universe$(1+w > 0)$, the cosmic
 expansion is singularity-free with the scale factor
$a(t)= a(t_m)\left[1+(1+w)(t - t_m)/t_m \right]^{\frac{2}{3(1+w)}}$.

Using Eq.(15), the energy density of the 
phantom universe $(t> t_m)$ is given by
%%%%%%%%%%%%%%%%%%%%%%%%%%%%%%%%%%%%%%%%%%%%%%%%%%%%%
\begin{equation}
\rho_x(t)\,=\,\frac{\rho(t_m)}{[-w+(1+w)t/t_m]^2}
\end{equation}

\begin{figure}[ht]  
\centering\includegraphics[height=7cm]{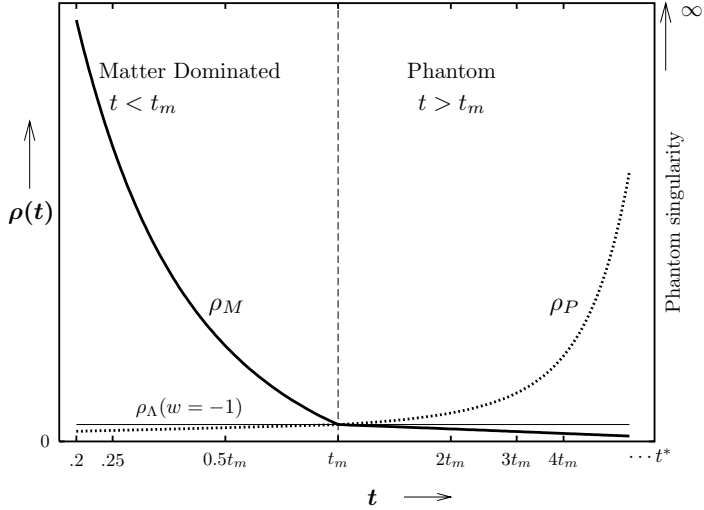}
\caption[]{\it Variation of the energy density in a phantom universe:
variation of the matter density is shown by the thick solid line,
variation of the vacuum energy density $\rho_\Lambda$ is 
shown by a thin line and the
variation of the phantom energy density by a dot line.}
\label{fig2}  
\end{figure}

Accordingly $\rho_x$ goes on increasing gradually,
 followed by steep rise to infinite value at 
$t^* = \frac{w}{1+w} t_m$. 
Therefore the phantom models have a finite life time ending in
a singularity. On  the other hand, 
$\rho_x$ scales down with time in the quintessence universe whereas
$\rho_\Lambda$ remains stationary. For $t< t_m$, 
the Hubble expansion is dominated by
matter density; 
accordingly $a\sim t^{2/3}$,\,\,,\,$\rho_m\sim t^{-2}$ but $\rho_x$ varies
independently as $t^{-2(1+w)}$ as shown in FIG.~\ref{fig2}.
%%%%%%%%%%%%%%%%%%%%%%%%%% Figure 2 %%%%%%%%%%%%%%%%%%%%%%%%%%%%%%%% 
%%%%%%%%%%%%%%%%%%%%%%%%%%%%%%%%%%%%%%%%%%%%%%%%%%%%%%%%%%%%%%%%%%%%%%% 

The expansion and density singularity 
in the phantom universe corresponds to  the curvature
singularity as the Ricci scalar $R_i^i$ also 
tends to infinity at this epoch. The total age $t^*$ of
the phantom universe depends upon the choice of $w$ 
and is larger for values of $w$ closer to
$-1$.  For example, $t^*= 21 t_m$ for $w= -1.05$ whereas $t^*= 6 t_m $ for 
$ w= -1.2$ 

\section{PRESENT AGE OF THE UNIVERSE }

The present age $t_0$ of the universe depends 
on the cosmological density parameters and the
equation of state parameter $w$ which determines the 
redshift $z_m$ at the transition epoch $t_m$
(at the end of matter dominated phase). We calculate $t_0$ in two steps. 
During the
matter-dominated phase, the Hubble expansion is given by
 Eq.(9). Integrating the first  term in
the binomial expansion of Eq.(9) over the redshift range 
from infinity to $z_m$, we calculate the age
of the universe from the beginning to 
the end of matter-dominated era as given by the expression
%%%%%%%%%%%%%%%%%%%%%%%%%%%%%%%%%%%%%%%%%%%%%%%%%%%%%%%%%%%
\begin{equation}
t_m \,=\,
 (H_0\sqrt{\Omega_m^0})^{-1}\left[ \frac{2}{3}(1+z_m)^{-3/2} \right]
\end{equation}

where $\Omega_m^0 = 0.27, \Omega_x^0 = 0.73$.

Using Eq.(15), the present age  $t_o$ is given by the equation
%%%%%%%%%%%%%%%%%%%%%%%%%%%%%%%%%%%%%%%%%%%%%%%%%%%%%
\begin{equation}
t_0\,\,=\,\left[1\,+\frac{(1+z_m)^{3(1+w)/2}-1}{1+w}\right]\times t_m
\end{equation}

Combining Eq.(17) and Eq.(18), the present
 age of the universe is calculated for a wide range of $ w $
as shown in the Table~\ref{table:age}.

\begin{table}
\begin{center}
\caption{\it  Age of the universe versus dark energy parameter $w$.
In the table $H_0^{-1} = 13.65$ Gyr.}
\vspace*{3mm}
\label{table:age}
\begin{tabular}{|c|c|c|c|c|}
\hline \hline
$w$ & $z_m$  & $t_m \; \; (H_0^{-1})$    &   $t_0 \;\; (H_0^{-1})$  & $t_0 \;\; (Gyr)$ \\  \hline
-0.66 & 0.644  & 0.602               & 1.112               & 15.18     \\  
-0.70 & 0.678  & 0.583               & 1.090               & 14.87     \\
-0.75 & 0.707  & 0.568               & 1.060               & 14.46     \\
-0.80 & 0.739  & 0.554               & 1.053               & 14.37     \\
-0.85 & 0.752  & 0.547               & 1.037               & 14.15     \\
-0.90 & 0.757  & 0.545               & 1.024               & 13.97     \\
-0.93 & 0.758  & 0.545               & 1.020               & 13.92     \\
-0.95 & 0.756  & 0.546               & 1.016               & 13.88     \\
-1.00 & 0.755  & 0.549               & 1.012               & 13.81     \\
-1.02 & 0.749  & 0.548               & 1.004               & 13.70     \\
-1.05 & 0.745  & 0.550               & 1.001               & 13.66     \\
-1.10 & 0.739  & 0.554               & 0.995               & 13.58     \\
-1.15 & 0.726  & 0.559               & 0.991               & 13.53     \\
-1.18 & 0.721  & 0.562               & 0.987               & 13.48     \\
-1.20 & 0.719  & 0.563               & 0.985               & 13.45     \\
-1.35 & 0.683  & 0.583               & 0.979               & 13.36     \\ 
-1.50 & 0.647  & 0.601               & 0.976               & 13.32     \\ \hline
\hline
\end{tabular}
\end{center}
\end{table}

\section{DUALITY  IN  QUINTESSENCE  AND  PHANTOM  BEHAVIOR}
\begin{figure}[ht]  
\centering\includegraphics[height=8cm]{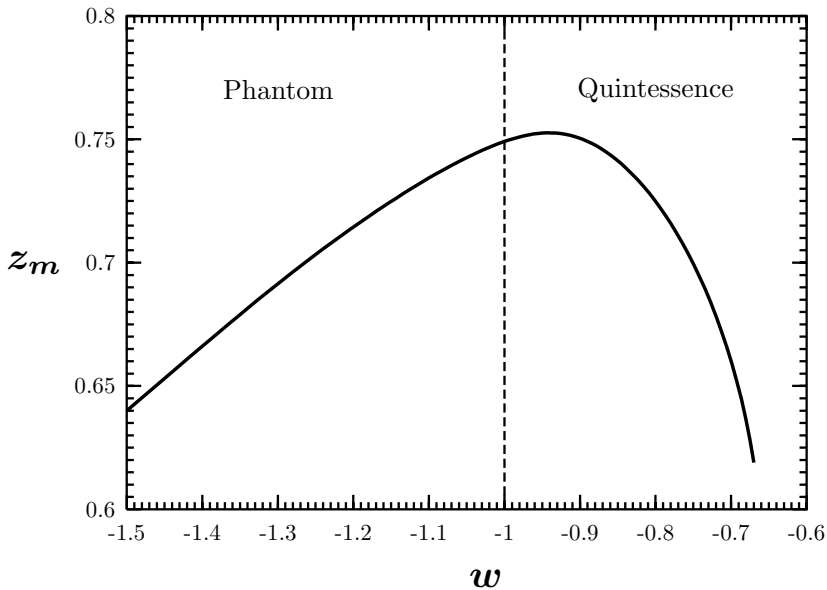}
\caption[]{\it Duality in the behavior of the phantom and the quintessence 
field is shown with respect to any chosen value of the transition redshift 
$z_m$. The peak value of $z_m$ lies in the quintessence field; 
the nearby $w$ in this field also show degeneracy with respect to $z_m$.}
\label{fig3}  
\end{figure}

We have investigated the correlation between the transition redshift $z_m$
(corresponding to the end of matter-dominated era) and the dark energy
 parameter  $w$ and found a sort of duality in the behavior of quintessence
fields (Q) and phantom fields (P) [see FIG.~\ref{fig3}]. 
For every value of $z_m$ in FIG.~\ref{fig3},
the corresponding parameter $w$ has,in general, 
two values, one lying in the range of
Q fields and the other in the range of P fields, both leading to cosmological
parameters (like the present age $t_0$ and the deceleration parameter $q_0$)
which seem to be compatible with the observational data. For example, if the
 transition from the decelerating phase to accelerating phase occurs at
$z_m = 0.739$, the corresponding dark energy parameter  may be 
either $w= \,-0.8$ (Quintessence field) or $w=\, -1.1$ (phantom field)
which yield  $14.37$ Gyr and $13.58$ Gyr respectively for the present age of 
the universe. Hence both the Q model and the P model seem to be
 compatible with the observational value~\cite{45} of $t_0$.
{\em This duality poses a
question whether the phantom fields really exist or they are merely 
 ghost fields which replicate the quintessence-like behavior for 
super-negative equation of state,  violating the dominant energy condition.}
It might explain the concordance of SNIa and galaxy cluster
 abundance observations in the 
extended $w\,-\,\Omega_m$ parameter space for $w<-1$ 
(phantom models) as shown by 
Caldwell et al~\cite{7}.

The above-mentioned duality is essentially different from  the form-invariance 
symmetry between standard cosmology and phantom cosmology pointed out 
independently by Dabrowski et al~\cite{24} and Chimento et al~\cite{55}.
 They have shown that 
in the case of a single component cosmological model, the scale factors of the 
standard and phantom models for a given energy density have reciprocal 
relationship 
with the equation of state parameter $w_{ph}\, = \,\, -w\,-2\,,(w>\,-1)$. This 
formalism seems inadequate to describe the evolution of the scale factor 
in dark energy models in conjunction  with pressure-free matter.

%%%%%%%%%%%%%%%%%%%%%%%%%% Figure 3 %%%%%%%%%%%%%%%%%%%%%%%%%%%%%%%% 
%%%%%%%%%%%%%%%%%%%%%%%%%%%%%%%%%%%%%%%%%%%%%%%%%%%%%%%%%%%%%%%%%%%%%%%%%% 

\section{CONSTRAINTS OVER PHANTOM COSMOLOGIES}

One of the greatest challenges in cosmology is to understand the nature
of the dark energy. Dark energy models are characterized by two parameters
$\Omega_x$ and $ w$. From the analysis of WMAP data~\cite{47},
 $\Omega_x = 0.73\pm 0.04$ is known up to high precision but $ w< -0.8$
(95 \% cl) leaves the field open to speculation whether the dark energy is 
phantom energy or quintessence energy. According to
 Melchiorri's analysis~\cite{13},
$-1.38< w < -0.82$. We have examined the possibility of  constraining the
 range of $w$ by comparison of the theoretically calculated age of the universe
 in Table 1 with the age derived from the observational data from WMAP, 
SDSS and data analysis of Tegmark et al~\cite{48}. 
By assuming $H_0\,=\,71_{-3.0}^{+4.0}$ and $\,t_0\,=\,13.7\pm0.2$ 
Gyr from WMAP data,
 we can find a narrow range $-1.18\,<\,w\,<\,-0.93$ for dark energy parameter 
and the corresponding range $-0.8\,<\,q_0\,<-0.52$ for the deceleration 
parameter (consistent with Kisilev's analysis~\cite{44}) 
but it would be more realistic
to allow for errors in the measurement of Hubble constant 
and take $ H_0\,=\,72\pm7.0\,
and\, t_0\,=\,13.0\pm1.5\, Gyr$~\cite{45} but 
this yields a wide range $-1.5\,<\,w\,<\,-0.75$ for 
variation of dark energy parameter and a very loose 
constraint on $w$ and $q_0$.

{\em However the degeneracy in the determination of of $w(z_m)$ for a chosen 
transition redshift may be used to constrain $w$ and discriminate between the
 quintessence and phantom dark energy.
Since the formation of the galactic clusters ceases with the end of 
matter dominated era at redshift $z_m$, the lowest redshift observations 
of the galactic clusters can indicate the maximum value of the redshift at 
which large scale structure formation in the universe would stop. In 
general there might be two values of
 $w$ corresponding to this particular value of $z_m$ out of which the one 
satisfying the WMAP range for the present age of 
the universe $(t_0 = 13.7\pm0.2)$, 
may be taken as the correct value of $w$ to determine 
the genuine  candidate of 
dark energy filling the universe.}

\section{CONCLUDING REMARKS}

The combined analysis of the latest cosmological observations provides a 
definite clue of 
the existence of dark energy in the universe but it is difficult to 
distinguish between 
the various forms of dark energy at present.
 So far as phantom energy is concerned, 
it is found to be compatible with SNe Ia observations 
and CMB anisotropy measurements 
but the violation of the `dominant
Energy condition' makes phantom models physically unstable; 
however, phantom models 
may be considered to be phenomenologically viable provided 
their age happens to be 
less than the time scale of the singularity. In case the instability 
occurs earlier 
and the dark energy decays into gravitons, as discussed
 by Carroll et al~\cite{14}, 
the universe might escape the ordeal of
`Rip-Off'~\cite{7} and phantom singularity. With the large number of SNe Ia 
observations expected from SNAP, LOSS and other surveys in the coming years, 
a clear picture of the dark energy profile is likely to emerge which would 
reveal whether $\Lambda$CDM, QCDM or PCDM is the 
correct cosmology of the universe.

\acknowledgments 
This work was done under the SERC Project, submitted to DST, India.
The author gratefully acknowledges the hospitality extended by 
W. I. Fine Theoretical Physics Insitute, University of Minnesota, 
useful discussions with
Keith Olive and valuable help of Vassilis Spanos and Roman Zwicky 
in the preparation of this manuscript.

\end{document}